\documentclass{article}
\usepackage{spconf}
\usepackage{graphicx} 
\usepackage[margin=3cm]{geometry}
\usepackage{amsmath}
\usepackage{hyperref} 
\usepackage{multirow}
\usepackage{booktabs}

\hypersetup{
     unicode=false,          
     pdftoolbar=true,        
     pdfmenubar=true,        
     pdffitwindow=false,     
     pdfstartview={FitH},    
     pdftitle={},    
     pdfauthor={The authors},     
     pdfsubject={},   
     pdfcreator={},   
     pdfproducer={}, 
     pdfkeywords={Blind color deconvolution}{histopathological images}{Bayesian modelling and inference}{variational Bayes}, 
     pdfnewwindow=true,      
     colorlinks=true,       
     linkcolor={blue},          
     citecolor=blue,        
     filecolor=blue,      
     urlcolor=blue           
 }

\usepackage{enumitem}
\setlist{nosep, leftmargin=14pt}

\newcommand{\etal}{\emph{et al.} }

\def\bn{{\mathbf{n}}}

\def\bx{{\mathbf{x}}}
\def\by{{\mathbf{y}}}

\def\b0{{\mathbf{0}}}

\def\bB{{\mathbf{B}}}

\def\bD{{\mathbf{D}}}

\def\bH{{\mathbf{H}}}
\def\bI{{\mathbf{I}}}

\def\balpha{{\boldsymbol{\alpha}}}

\def\btheta{{\boldsymbol{\theta}}}

\newcommand{\p}{{\mathrm p}}

\newcommand{\bTV}{{\mathrm {TV}}}

\def\bn{{\mathbf n}}

\def\bx{{\mathbf x}}
\def\by{{\mathbf y}}

\def\bB{{\mathbf B}}

\def\bD{{\mathbf D}}

\def\bH{{\mathbf H}}
\def\bI{{\mathbf I}}

\def\balpha{{\boldsymbol{\alpha}}}

\def\btheta{{\boldsymbol{\theta}}}

\title{TV-based Deep 3D Self Super-Resolution for fMRI}
\name{
\begin{tabular}{c}
Fernando Pérez-Bueno$^{a,}$\sthanks{Corresponding author. **Senior authors contributed equally \\The code will be made available at https://github.com/zalteck},
 Hongwei B. Li$^{b}$, Matthew S. Rosen$^{b}$, Shahin Nasr$^{b}$, \\César Caballero-Gaudes$^{a,c,**}$, Juan E. Iglesias$^{b,e,d,**}$
 \end{tabular}
 }
\address{$^{a}$ Basque Center on Cognition, Brain, and Language (BCBL), Spain.\\ 
$^{b}$ Athinoula A. Martinos Center for Biomedical Imaging, Harvard Medical School, USA\\
$^{c}$ Ikerbasque, Basque Foundation for Science, Bilbao, Spain\\
$^{d}$ CSAIL, Massachusetts Institute of Technology (MIT), USA\\
$^{e}$ Center for Medical Image Computing University College London (UCL), U.K 
}
\begin{document}
%
\maketitle
\begin{abstract}
While functional Magnetic Resonance Imaging (fMRI) offers valuable insights into cognitive processes, its inherent spatial limitations pose challenges for detailed analysis of the fine-grained functional architecture of the brain. More specifically, MRI scanner and sequence specifications impose a trade-off between temporal resolution, spatial resolution, signal-to-noise ratio, and scan time. Deep Learning (DL) Super-Resolution (SR) methods have emerged as a promising solution to enhance fMRI resolution, generating high-resolution (HR) images from low-resolution (LR) images typically acquired with lower scanning times. However, most existing SR approaches depend on supervised DL techniques, which require training ground truth (GT) HR data, which is often difficult to acquire and simultaneously sets a bound for how far SR can go.
In this paper, we introduce a novel self-supervised DL SR model that combines a DL network with an analytical approach and Total Variation (TV) regularization. Our method eliminates the need for external GT images, achieving competitive performance compared to supervised DL techniques and preserving the functional maps.
\end{abstract}
\section{Introduction}


\begin{figure*}
    \centering
    \footnotesize
    \vspace{-5pt}
    \includegraphics[width=0.85\textwidth]{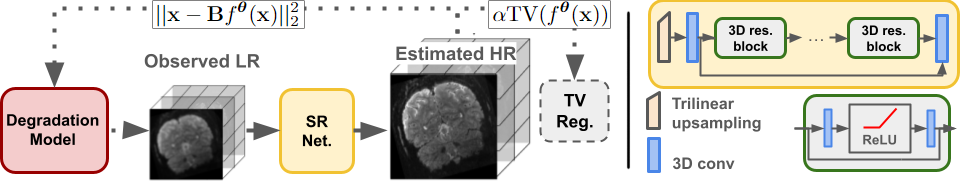}
    \vspace{-10pt}
    \caption{Overview of the proposed self SR model. The observed LR image is fed to the SR Network to produce an HR estimation. During training, the HR output is used to calculate the TV regularization and downsampled to be compared to the observed input. 
    }
    \label{fig:model_overview}
    \vspace{-10pt}
\end{figure*}

fMRI provides a non-invasive window for observing brain function \emph{in vivo}. However, the spatial resolution of fMRI scans is significantly lower than other MRI approaches, as it is limited by the trade-off with scan time, spatial resolution, and signal-to-noise ratio~\cite{VuTradeoff}. 
The use of ultra-high-field scanners enables the capture of images at the mesoscale organization of the brain, yet the acquisition of higher spatial resolution remains challenging due to the increased scan times required to compensate for the low signal and contrast-to-noise ratio.
Specifically, limits in spatial resolution hampers the ability to study fine-scale neural processes involved in sensory and cognitive processing, e.g. at the level of cortical layers or small subcortical regions. 

 Super Resolution (SR) methods, especially Deep Learning (DL) based, have arisen as a possible solution to circumvent the limitations of fMRI~\cite{Bran_Agnostic}. However, most recent SR approaches for fMRI use a supervised approach that relies on pairs of LR and HR images ~\cite{Lugmayr_RealWorldSR}. While this approach can effectively yield SR fMRI images, it has two major drawbacks: (i) It requires access to adequate GT images for supervised learning, and (ii) it cannot surpass current fMRI limitations, as the resolution is limited to that of the available GT images.

This paper presents a novel approach to overcoming these limitations through the development of a self-supervised 3D DL SR method specifically designed for fMRI data. Our model assumes a degradation model and a Total Variation (TV) prior for the HR images and trains a Convolutional neural network (CNN) using only the observed data. Our method does not require HR GT for training and can double the resolution of the input.

\subsection{Related work}

Very few works have explored the SR of fMRI and its impact on functional maps.
One of the pioneer works was the 2D patch-based approach in~\cite{Kornprobst2003_SRfmri}, which combined an adapted fMRI acquisition with in-plane model-based SR with a Huber regularization. This work, and its extension in~\cite{Peeters2004}, facilitate the acquisition of fMRI by reducing the slice thickness with SR. 

The lack of matching fMRI training data, which should have the same contrast, field of view, sequence, and even possible pathology as the test data to process, makes it difficult to adapt supervised SR methods for every need. This domain shift problem motivated the works in~\cite{Ota2022} by using a Generative Adversarial Network (GAN) to produce synthetic HR fMRI using T2*-weighted HR images as a reference, and in~\cite{Bran_Agnostic} by training a 3D-CNN on resting state (RS) images and testing on visual-related tasks.

SR methods are explored more widely in other MRI modalities. The supervised approach with paired LR (synthetic) and HR training data is the most common and has been explored using diverse methods such as Compressive Sensing~\cite{Li2019}, Random Forest~\cite{Jog14_SupervisedSR}, or non-parametric patch-based~\cite{Rousseau_patchbased}. Supervised CNNs have been applied to the field of MRI SR in combination with spline interpolation~\cite{Pham2019} or nearest neighbors~\cite{Shi2019}. The different resolutions between planes on anisotropic MRI have also been used as self-similarity~\cite{Zhao2018, SMORE_SelfSR21} to reconstruct thick slice MRI.

However, despite recent advances in DL, single-image analytical interpolation methods are still commonly used in MRI~\cite{SMORE_SelfSR21}. They use an observation model for the LR image and regularization terms for the HR image such as TV~\cite{Tourbier2015, Feng2015}, or low-rank matrices~\cite{Feng2015}, but require optimization for each new image.

\vspace{-10pt}
\section{Method}
We propose a DL SR CNN, trained in a self-supervised loss building on the classical analytical approach~\cite{BCDnet}.
Let us denote $\by$ the vectorized HR 3D image (unavailable), and $\bx$ the vectorized LR image (observed).

The relation between the HR and LR images is defined by $\bx= \bD\bH\by + \bn = \bB\by + \bn$
where $\bn$ is assumed to be independent Gaussian noise, and $\bB = \bD\bH $ is a downsampling operator that combines decimation ($\bD$) and blurring ($\bH$). Then, the observation model can be expressed using a normal distribution.
\vspace{-2pt}
\begin{equation}
\vspace{-2pt}
    \p(\bx | \by; \lambda^2)= \mathcal{N}(\bx|\bB\by, \lambda^2\bI)
\end{equation}
where $\lambda^2$ is the noise variance. Although the actual downsampling operator is also unknown, we assume it is linear, which is a common choice in SR~\cite{Bran_Agnostic, Pham2019}.

A TV prior is included for the unknown HR image $\by$ to promote a sharp image. This prior is a common choice~\cite{Tourbier2015, Feng2015} in analytical approaches. It is easy to calculate and has good noise-reduction properties while preserving edges in the image. Furthermore, it has been proven to work well in combination with deep priors\cite{SelfDeblur}. We use $\p(\by|\balpha) \propto  \exp \left[ -\alpha \bTV(\by) \right]$
with $\alpha>0$ controlling the image smoothness. The TV function is the isotropic 3D norm of the first-order differences for each axis at each image voxel.

\begin{table*}[ht]
    \centering
    \footnotesize
    \caption{Comparison with interpolation methods. No GT is required, all methods run on the validation set only. 
    }
    \begin{tabular}{cccccccccccc}
         Input &  & \multicolumn{2}{c}{Trilinear} & \multicolumn{2}{c}{Nearest} & \multicolumn{2}{c}{3D-spline} & \multicolumn{2}{c}{Proposed} \\
         (mm)& Factor& PSNR & SSIM& PSNR & SSIM& PSNR & SSIM& PSNR & SSIM\\
         \midrule
         1.875&$\times$1.25& 78.16 & 0.9446 & 74.92 & 0.9351& 81.01 & 0.9571& \textbf{83.93} & \textbf{0.9673}\\
         2.25&$\times$1.5& 75.07 & 0.9265 & 70.68 & 0.9077& 75.57 & 0.9334&\textbf{79.76} & \textbf{0.9497}\\
         2.62&$\times$1.75& 73.54 & 0.9169 & 66.90 & 0.8712& 71.60 & 0.9073& \textbf{76.50} & \textbf{0.9365}\\
         3&$\times$2& 69.59 & 0.8855 & 68.47 & 0.8874& 67.42 & 0.8729&\textbf{73.53}&\textbf{0.9218}\\ 
         \bottomrule
    \end{tabular}
    \label{tab:Interpolation_OpNeuro}
    \vspace{-10pt}
\end{table*}

\begin{table*}[ht]
    \centering
    \footnotesize
    \caption{Comparison with DL methods. Separated Training/Validation sets. Target resolution $1.5mm$}
    \begin{tabular}{cccccccccccc}
         input&  & \multicolumn{2}{c}{Agnostic-Li} & \multicolumn{2}{c}{Super-Li} & \multicolumn{2}{c}{Super-MSE} & \multicolumn{2}{c}{DIP}& \multicolumn{2}{c}{Proposed} \\
         (mm)& Factor&PSNR & SSIM& PSNR & SSIM& PSNR & SSIM& PSNR & SSIM& PSNR & SSIM\\
         \midrule
         1.857&$\times$1.25& 84.09  & 0.9707  & 85.91 & \textbf{0.9733} & \textbf{86.21} & 0.9731& 84.54 & 0.9648 & 85.82 & 0.9713\\
         2.25&$\times$1.5& 81.32 & 0.9590  & 80.46 & 0.9572 & \textbf{82.46} & \textbf{0.9607} & 79.85 & 0.9503 & 80.04 & 0.9546\\
         2.62&$\times$1.75& 77.43 & 0.9426 & 78.58 & \textbf{0.9468} & \textbf{78.66} & 0.9460 & 75.74 & 0.9273 & 76.72 & 0.9368\\
         3&$\times$2& 73.78 & 0.9204 & \textbf{76.32} & \textbf{0.9375} & 76.27 & 0.9332 & 73.17 & 0.9104 & 73.45 & 0.9192\\
         \bottomrule
    \end{tabular}

    \label{tab:DL_OpNeuro}
    \vspace{-10pt}
\end{table*}

\vspace{-7pt}
\subsection{Deep inference}
 

Instead of classical variational inference, which requires bounding the TV prior and re-training the model for each new image~\cite{PEREZBUENO_TV}, we use a combination of analytical and DL modeling~\cite{BCDnet}, where a CNN $f^\btheta(\cdot)$ predicts a MAP estimate for $\by$ (see figure~\ref{fig:model_overview}).

This choice simplifies the Evidence Lower Bound (ELBO) into the classical CNN loss function with fidelity and regularization terms~\cite{BCDnet}. The noise in the observation model is assumed constant for all images in the dataset. Hence, both terms are weighted using the confidence parameter $\alpha$ on the TV prior:
\begin{equation}
    \mathcal{L}=||\bx - \bB f^\btheta(\bx)||^2_2 + \alpha\bTV(f^\btheta(\bx))
    \label{eq:Loss}
\end{equation}


The first term (``fidelity") encourages the SR network $f^\btheta$ to produce an output that, when downsampled, is as close as possible to the observed LR image. The second term (``TV regularizer") promotes a sharp HR image where the edges are preserved. 

This leads to a self-supervised model that requires no HR GT image for training. When $\alpha=0$ it becomes an unconstrained Deep Image Prior (DIP) approach~\cite{Ulyanov2020} where the only requisite for an estimated HR image is to be the output of a CNN.

\vspace{-5pt}
\subsection{Network architecture and training details}

Following~\cite{Bran_Agnostic} we use trilinear interpolation followed by a 3D fully CNN with ten dense residual layers, kernel size 3, and zero padding. The interpolation enables us to use the same architecture, irrespective of the upscaling factor used. The convolutional layers and residual connections learn local features that allow us to recover fine details in the upscaled image. 
We use an Adam optimizer with an initial learning rate of $10^{-3}$, halved on the plateau after 5 epochs. 

We train the SR network for a fixed upsampling factor $f \in \{1.25, 1.5, 1.75, 2\}$. The network can be trained using a single image (non-amortized) or with a dataset of images (amortized). While the single-image training might offer a higher structural fidelity, the amortized approach significantly reduces the testing computational cost and might improve SR performance~\cite{Pham2019}. Therefore, our network is trained with random images from the time series of as many subjects and runs are available in a given dataset.



\vspace{-10pt}
\begin{figure*}[ht]
\vspace{-10pt}
    \centering
    \def\mysize{0.28}
    \begin{tabular}{ccc}
    a) Original & b)$f= \times1.25$ & c) $f= \times 2$ \\
        \includegraphics[width=\mysize\textwidth]{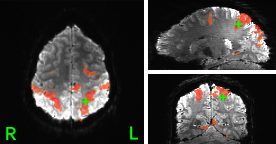}&
        \includegraphics[width=\mysize\textwidth]{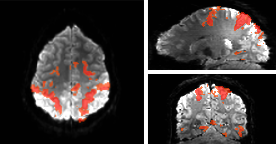}&
        \includegraphics[width=\mysize\textwidth]{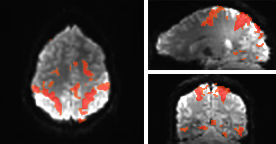}
        \\
         \includegraphics[width=\mysize\textwidth]{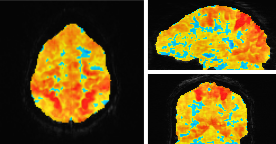}&
        \includegraphics[width=\mysize\textwidth]{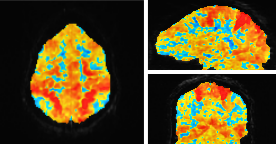}&
        \includegraphics[width=\mysize\textwidth]{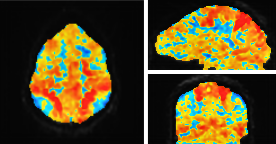}
        \\
    \end{tabular}
    \vspace{-5pt}
    \caption{Single-subject seed correlation maps. Top: Thresholded maps ($r \geq 0.5$) showing a clear pattern of the sensorimotor network with clusters in bilateral motor and somatosensory cortices, and supplementary motor areas. Bottom: Unthresholded maps. a) Real observed image $1.5mm$. Super-resolved image at $1.5mm$ from an input of isotropic voxels size b) $1.875mm$ $(f=1.25)$ and c) $3mm$ $(f=2)$.}
    \label{fig:functional_resting_maps}
    \vspace{-10pt}
\end{figure*}

\vspace{-2pt}
\section{Experiments and Results}
\vspace{-5pt}
\subsection{Data: Gorgolewski Resting State dataset}
\label{sec:data}
This publicly available dataset~\cite{OpNeuro_Gorgo} contains whole-brain T2*-weighted fMRI scans acquired at 7T with a gradient-recalled-echo echo planar imaging sequence during 15 min while subjects remained at rest with eyes-open (FOV= $192\times192mm^2$ (R-L; A-P), matrix size=$128\times128$, 70 axial oblique slices with 1.5 $mm$ isotropic voxel size, TR=3.0 s, TE=17 ms, FA=70°, Partial Fourier 6/8, GRAPPA=3 with 36 reference lines). For further details, please see~\cite{OpNeuro_Gorgo}. We used data from 12 participants, avoiding those with reported acquisition issues. The first eight subjects were used for training and the other four for validation.

\vspace{-5pt}
\subsection{Image quality comparison}
\vspace{-2pt}
We compare our proposed method with standard interpolation approaches (trilinear, nearest neighbor, and cubic spline implemented in nilearn) and the following DL-SR methods: the resolution agnostic method proposed by Li~\etal
\cite{Bran_Agnostic} (Agnostic-Li), the fixed-factor supervised version of the same method (Super-Li), our base network~\cite{Pham2019} using supervised training (Super-MSE), and the unconstrained DIP approach~\cite{Ulyanov2020}.
For this comparison, we generated synthetic LR observations that were then upsampled to the resolution of the original images ($1.5mm$). The value of the hyperparameter $\alpha$ in eq.~\eqref{eq:Loss} was experimentally set to $0.01$ after a grid search in the interval $\{0,0.1\}$. Peak Signal-to-Noise Ratio (PSNR) and Structural Similarity (SSIM) were calculated between the original (GT) and reconstructed images using validation data.

Table~\ref{tab:Interpolation_OpNeuro} shows the comparison with the interpolation methods, where only the validation data is used and no GT is required. Table~\ref{tab:DL_OpNeuro} reports the results of the DL approaches, where the methods are trained and validated in separated subjects. The training subset is the same for all methods, but the supervised approaches have access to the training GT while DIP and the proposed method only have access to the LR observations. The proposed method outperforms the interpolation methods and is competitive against the supervised DL approaches despite training only with the LR images. The proposed method also outperforms the unconstrained DIP approach, showing that the addition of the TV prior can improve the performance to up to 1.28 dB for an upscaling factor of $\times1.25$. 
The larger training set used in table~\ref{tab:DL_OpNeuro} is beneficial for the proposed method, which performs better than when trained only with the validation set. 


\vspace{-5pt}
\subsection{Functional analysis}
In this section, we evaluate the effect of the proposed SR algorithm on seed-based correlation analyses to assess functional connectivity in single datasets. A minimal data preprocessing pipeline was applied to a single subject's original and reconstructed images in the testing set, including volume realignment, nuisance regression of up to 4th-order Legendre polynomials, motion parameters, and their derivatives, band-pass filtering between $0.005-2$ Hz, censoring scans with excessive motion (Euclidean norm of motion derivatives $\geq 0.3$), and spatial smoothing with $3mm$ FWHM Gaussian kernel\footnote{AFNI commands are included in the available code.}. Seed correlation maps were calculated using AFNI's \textit{@InstaCorr} plugin from a seed of radius of $3 mm$. 

The original and reconstructed brain masks from \textit{3dAutomask} had a Jaccard similarity index of $0.9769$ and $0.9684$ for factors $\times1.25$ and $\times2$, respectively. Figure~\ref{fig:functional_resting_maps} displays the corresponding seed correlation maps from a seed located in the precentral gyrus of the primary motor cortex. Both SR reconstructed maps exhibit a high spatial similarity, although the larger upsampling factor ($\times2$) produces a slightly more blurred map. The edge-preserving effect of the TV prior can be seen more clearly in the unthresholded maps shown in the bottom row of Figure~\ref{fig:functional_resting_maps} as the factor increases, where the edges of the temporally-correlated clusters are more clearly delimited.


For a quantitative comparison, we binarize the thresholded functional maps and consider those obtained from the original image as GT. The accuracy and false discovery rate (FDR) are presented in Table~\ref{tab:Func_quantitative}.
The resulting maps achieve an average accuracy of 0.9694 and 0.9129 and a false discovery rate of $0.0064\%$ and $0.0120\%$ for factors $\times1.25$ and $\times2$, respectively. This demonstrates the ability of the proposed method to preserve functional analysis. 



\begin{table}[ht]
\footnotesize
\vspace{-10pt}
    \centering
        \caption{Accuracy and FDR(\%) for the functional maps obtained with the reconstructed images.}
        \setlength{\tabcolsep}{4pt} 
    \begin{tabular}{ccccc}
         & \multicolumn{2}{c}{$f = \times 1.25$} & \multicolumn{2}{c}{$f = \times 2$}\\
        RS Network & Acc. & FDR  & Acc. & FDR   \\
        \midrule
        Sensory Motor & 0.9690 & 0.0099 & 0.9598 & 0.0128\\
        Default Mode & 0.9522 & 0.0055 & 0.7916 & 0.0209\\
        Visual & 0.9869 & 0.0038 & 0.9873 & 0.0024 \\
        \bottomrule
    \end{tabular}

    \label{tab:Func_quantitative}
    \vspace{-15pt}
\end{table}

\vspace{-2pt}
\section{Discussion and Conclusion}

This study introduces a novel method for 3D self super-resolution of fMRI images that integrates DL with a TV prior and does not require HR GT data for training. Our approach demonstrates competitive performance compared with supervised DL methods, effectively enhancing the spatial resolution of fMRI data while preserving the integrity of RS functional analyses. The use of a TV prior enhances the output of the CNN and promotes that the super-resolved images retain the sharp edges while reducing the noise. These results suggest that our method offers a promising alternative to enhance fMRI spatial resolution without compromising functional information. 

\vspace{-5pt}
\section{Compliance with ethical standards}
\label{sec:ethics}
\vspace{-5pt}

This research study was conducted retrospectively using human subject data made available in open access~\cite{OpNeuro_Gorgo}. Ethical approval was not required as confirmed by the license attached with the data.

\vspace{-10pt}
\section{Acknowledgments}
\label{sec:acknowledgments}
\vspace{-5pt}
This research is supported by JDC2022-048784-I, funded by MCIN/ AEI/ 10.13039/ 501100011033 and the European Union “NextGenerationEU” / PRTR, by the Basque Government (BERC 2022-2025 program), by the Spanish State Research Agency (BCBL Severo Ochoa excellence accreditation CEX2020-001010/ AEI/
10.13039/ 501100011033),
and by NIH (RF1MH123195, R01AG070988,
UM1MH130981, RF1AG080371)
\vspace{-7pt}

\bibliographystyle{IEEEbib}
\bibliography{biblio_reduced}

\end{document}